\documentclass{article}
\usepackage{amsfonts,amssymb, amsmath}
\usepackage[english]{babel}

\def\bq{ \begin{equation} }
\def\eq{ \end{equation} }
\def\ben{ \begin{eqnarray} }
\def\en{ \end{eqnarray} }

\newtheorem{prop}{Proposition}
\begin{document}


\title{Comment on "The separation of variables and bifurcations of first integrals in one problem of D.N.Goryachev"\, by Pavel E. Ryabov (Archive:1102.2588v1)}
\author{A.V. Tsiganov \\
\it\small
St.Petersburg State University, St.Petersburg, Russia\\
\it\small e--mail:  andrey.tsiganov@gmail.com}

\date{}
\maketitle

\begin{abstract}
We prove that in the Ryabov paper an application of  the geometric Kharlamov method 
to  the Goryachev system  yields  noncommutative "new variables of separation"  
instead of the standard canonical variables of separation.
\end{abstract}

In order to describe Goryachev system  we will use
the angular momentum vector $M=(M_1,M_2,M_3)$ and the Poisson vector $\alpha=(\alpha_1,\alpha_2,\alpha_3)$ in a moving frame of coordinates attached to the principal axes of inertia. The Poisson brackets between these variables
\begin{equation}\label{e3}
\,\qquad \bigl\{M_i\,,M_j\,\bigr\}=\varepsilon_{ijk}M_k\,, \qquad
\bigl\{M_i\,,\alpha_j\,\bigr\}=\varepsilon_{ijk}\alpha_k \,, \qquad
\bigl\{\alpha_i\,,\alpha_j\,\bigr\}=0\,,
\end{equation}
may be associated to the Lie-Poisson brackets on the algebra $e^*(3)$.  Let us consider dynamical system defined by the following integrals of motion
 \ben
H=\dfrac{1}{2}(M_1^2+M_2^2)+M_3^2+\dfrac{1}{2}c(\alpha_1^2-\alpha_2^2)+\dfrac{b}{\alpha_3^2}\,,\qquad b,c\in \mathbf R\,,\nonumber\\
\label{chap-hg}\\
K=\left(M_1^2+M_2^2+\dfrac{b}{\alpha_3^2}\right)^2+2c\alpha_3^2(M_1^2-M_2^2)+c^2\alpha_3^4\,,
\nonumber
\en
Because
\[
\{H,K\}=8c \alpha_3M_1M_2(\alpha_1M_1+\alpha_2M_2+\alpha_3M_3)
\]
it is  integrable system on the four-dimensional symplectic leaves of $e^*(3)$ on
the zero level of the Casimir function
\bq\label{zlevel}
C=\alpha_1M_1+\alpha_2M_2+\alpha_3M_3=0\,.
\eq
At  $b=0$ this system and the corresponding variables of separation have  been investigated by Chaplygin  \cite{ch03}.  Some generalizations of the Chaplygin results  are discussed in \cite{bm05}.

Singular term has been added by Goryachev in \cite{gor16}.   In  \cite{ts10} we proved that the  Chaplygin variables remain  variables of separation for the Goryachev case at $b\neq 0$.

Both these  systems are subsystems of the so-called Kowalewski-Chaplygin-Goryaschev \cite{kuzts} with linear in variables additional terms in the Hamilton function. In \cite{ts02} we get complex variables of separation for this generic system using Lax matrices and  $r$-matrix theory.  The overview of the various applications of the  separation of variables method and discussion of this generic system may be found in \cite{bm03}.

 In \cite{r}  author declared that completely new variables of separation  for the Goryachev system may be found by using so-called geometric Kharlamov method \cite{h}.  According to \cite{r}  at  $c=1$ and  $\alpha_1^2+\alpha_2^2+\alpha_3^2=1$ new variables of separation  $u_1, u_2$ are roots of the following polynomial
 \bq\label{def-r}
 zu^2-2bu+(2b\xi-kz)=0\,,\qquad\mbox{где}\qquad z=\alpha_3^2\,,\quad \xi=M_1^2+M_2^2+\dfrac{b}{\alpha_3^2}\,,
 \eq
 where  $k$ is an integral  $K$.
 \begin{prop}
 The roots $u_1, u_2$ of the polynomial (\ref{def-r}) do not commute
\[\{u_1,u_2\}\neq 0\,,\]
with respect to the initial Poisson bracket (\ref{e3}) even on the zero-level of the Casimir function  $C$ (\ref{zlevel}).
 \end{prop}
According to  \cite{jac36},  variables of separation are canonical variables with respect to the Poisson brackets.
It means that variables $u_{1,2}$ can not be variables of separation and it is a crucial error in \cite{r}.

So, the geometric Kharlamov method \cite{h}  yields  some noncommutative variables instead of the canonical variables of separation for the Goryachev systems.  We try to understand what this means by using Chaplygin variables \cite{ch03}
\[
q_{1,2}=\dfrac{M_1^2+M_2^2\pm h}{\alpha_3^2}\,,\qquad h^2=(M_1^2-M_2^2+\alpha_3^2)^2+4M_1^2M_2^2\,.
\]
On the zero-level of the  Casimir function $C$ (\ref{zlevel}) variables  $q_{1,2}$ commute
\[\{q_1,q_2\}=0\,,\]
with respect to the Poisson bracket (\ref{e3}) according to the standard definition of the  variables of separation   \cite{jac36}. Using canonical  Chaplygin variables  $q_{1,2}$ and the corresponding conjugated momenta  $p_{1,2}$\cite{ts10,ts11} 
\[\{q_i,q_j\}=\{p_i,p_j\}=0\,,\qquad \{q_i,p_j\}=\delta_{ij}\,,\]
we can easily rewrite noncommutative roots of the polynomial  (\ref{def-r}) at the following form
\[
u_1=8(q_1^2-1)p_1^2-2q_1+2H\,,\qquad u_2=8(q_2^2-1)p_2^2-2q_2+2H\,.
\]
According to \cite{jac36},  any function $s_i(p_i,q_i)$ on a pair of canonical variables is variables of separation as well. Thus, using auxiliary variables $u_{1,2}$  we can obtain a desired set of the standard variables of separation
\[
s_1(q_1,p_1)=u_1-2H\,,\qquad s_2(q_2,p_2)=u_2-2H\,,
\]
which are equivalent to the Chaplygin variables.  It is a remarkable well-known shift
of auxiliary  variables $u_{1,2}$, which   Kowalevski used in \cite{kow89}  in order to get canonical
variables of separation $s_{1,2}$ in her case.

Summing up, in contrast with the methods of bi-hamiltonian geometry \cite{ts10,ts11},
the geometric Kharlamov method \cite{h} has to be supplemented by some additional tools
in order to get canonical variables of separation.

\end{document}